\documentstyle[preprint,aps]{revtex}
\begin{document}
\title{Transverse Isotropy in Identical Particle Scattering}
\author{L.F.~Canto$^1$, R.~Donangelo$^1$ and M.S.~Hussein$^2$}
\address{ 
$^1$Instituto de F\'\i sica, Universidade Federal do Rio de Janeiro,\\
C.P. 68528, 21945-970 Rio de Janeiro, Brazil\\
$^2$Instituto de F\'\i sica, Universidade de S\~ao Paulo,\\
C.P. 66318, 05389-970 S\~ao Paulo, Brazil}

\maketitle
\bigskip
PACS: 03.65.Nk

\begin{abstract}
It is pointed out that the cross section for the scattering of identical
charged bosons is isotropic over a broad angular range around 90$^o$
when the Sommerfeld parameter has a critical value, which depends
exclusively on the spin of the particle.
A discussion of systems where this phenomenon can be observed is presented.
\end{abstract}
\bigskip

The scattering of identical particles is a routine exercise in
quantum mechanics and its discussion can be found in
most text books on the subject \cite{LL}. However, in recent years,
the rapid oscillation seen in the angular distribution of the
Mott elastic scattering of identical charged particles such as
nuclei, was utilized to test models such as QCD \cite{HLPB,Villari} and to
discuss small deviations from the Coulomb force law owing to
QED-related corrections such as vacuum polarization \cite{Lynch}.

In the present work we point out a hitherto unknown feature of
the Mott scattering cross-section for bosons, namely an almost
isotropic angular distribution over a very wide angular range
when the Sommerfeld parameter attains a critical value determined
entirely by the spin, $s$, of the particles viz,
$\eta_C = (3 s +2)^{1/2}$.
For the purpose of completeness we first give a short account of 
the theory of scattering of identical particles. We then turn to
the derivation of $\eta_C$ and apply to several boson-boson 
scattering systems. We also briefly discuss the fermion-fermion case.

If in a scattering process the projectile and target particles are
identical, when one of them reaches a detector the experiment cannot tell if
this particle is the projectile or the target. On the other hand, momentum
conservation guarantees that whenever a particle emerges in one direction,
the other emerges in the opposite orientation, in the CM frame of reference.
Therefore, the amplitude for scattering at the orientations ${\bf r}$ and
$- {\bf r}$ will be mixed in some way. In Quantum mechanics, the total wave
function for pairs of identical particles with integer spins, i.e. two
bosons, must be symmetric with respect to the exchange of these particles,
while in the case of particles with half-integer spin, i.e. two fermions, it
must be anti-symmetric\label{Landau}.

The situation is simple for spinless bosons or in collisions where the
projectile and the target are polarized so that their spins are aligned. In
such cases, the wave function in the spin space is always symmetric and
projectile-target exchange reduces to reflection of the relative vector
position ${\bf r}$. For spherically symmetric potentials, there is axial
symmetry and space reflection corresponds to the transformation $\theta
\rightarrow \pi -\theta $. The elastic cross section is then given by 
\begin{equation}
\sigma _{\pm }(\theta )=\left| f(\theta )\pm f(\pi -\theta )\right| ^{2}\,,
\label{bf}
\end{equation}
where $\ f(\theta )$ is the scattering amplitude for discernible particles
of the same mass under the same potential $V(r)$.
The $+(-)$ sign in eq.(\ref {bf}) applies when the particles involved are
bosons (fermions).

\noindent

Eq.(\ref{bf}) may be rewritten in the form 
\begin{equation}
\sigma _{\pm }(\theta )=\sigma _{inc}(\theta )\pm \sigma _{int}(\theta )\,,
\label{split}
\end{equation}
with 
\begin{equation}
\sigma _{inc}(\theta )=\left| f(\theta )\right| ^{2}\,+\,\left| f(\pi
-\theta )\right| ^{2}  \label{inc}
\end{equation}
and 
\begin{equation}
\sigma _{int}(\theta )\,=2{\rm Re}\left\{ f^{\ast }(\theta )\,f(\pi -\theta
)\right\} \,.  \label{int}
\end{equation}

\noindent The first term in eq.(\ref{split}) is the incoherent sum of the
contributions to the cross sections arising from projectile and target, if
they were distinguishable. While this term is independent of the particle
statistics, the sign of the second term is responsible for the difference in
the expressions for the cross section of bosons and fermions. This
interference term has no classical analogue.

The situation is more complicated in the case of unpolarized spins.
In this case, the cross section mixes different parities as the spins
couple to produce symmetric or anti-symmetric states in the spin space.
However, taking the proper average over spin orientations one obtains the
simple formula 
\begin{equation}
\sigma _{\pm }(\theta )=\sigma _{inc}(\theta )\pm \frac{\sigma _{int}(\theta
)}{2s+1}\,,  \label{unpolarized}
\end{equation}
where $s$ is the spin of the particle in units of $\hbar$. 
Eq.(\ref{unpolarized}) indicates that the relevance of the interference term
decreases with the spin value, vanishing in the classical limit 
$s\rightarrow \infty $.

The cross sections of eq.(\ref{unpolarized}) are symmetric with respect to $%
\theta =90^o$ and their particular shape depends on several factors
such as the statistics of the colliding particles, their interaction and the
bombarding energy. A particularly interesting situation is the Coulomb
scattering of bosons, where $\sigma _{+}$ is known as the Mott cross section
and denoted $\sigma _{Mott}$. In this case, we have the analytical
expressions 
\begin{equation}
\sigma _{inc}(\theta )=\frac{a^{2}}{4}\,\left[ \frac{1}{\sin ^{4}\left(
\theta /2\right) }+\frac{1}{\cos ^{4}\left( \theta /2\right) }\right]
\label{inc-coul}
\end{equation}
and 
\begin{equation}
\sigma _{int}(\theta )=\frac{a^{2}}{4}\,\left[ \frac{2}{\sin ^{2}\left(
\theta /2\right) \,\cos ^{2}\left( \theta /2\right) }\,\,\cos \left( 2\eta
\ln (\tan ^{-1}(\theta /2))\right) \right] \,,  \label{int-coul}
\end{equation}
where $\eta $ is the Sommerfeld parameter, 
\begin{equation}
\eta =\frac{q^{2}}{\hbar v}  \label{eta}
\end{equation}
and $a$ is half the distance of closest approach in a head-on collision, 
\begin{equation}
a=\frac{q^{2}}{2E}\,.  \label{a}
\end{equation}
Above, $q$ is the charge of each of the two collision partners, $E$ is the
bombarding energy in the center of mass (CM)\ reference frame and $v$ is the
corresponding velocity of the relative motion. In the present case, 
$\sigma _{inc}$ exhibits a minimum at $\theta =90^o$, with the value 
$\sigma _{inc}\left( \theta =90^o\right) =2a^2$, with an
energy-independent shape. On the other hand, $\sigma _{int}$ has always a
maximum at this angle, with the same value $2a^{2}$. However, its shape
depends on the collision energy through the Sommerfeld parameter $\eta $.
The behavior of $\sigma _{Mott}$ in the vicinity of $\theta =90^o$
results from a competition between these two opposing trends. For small 
$\eta $ values, $\sigma _{int}$ is a slowly varying function of $\theta $.
The shape of $\sigma _{Mott}$ is then dominated by that of $\sigma _{inc}$
and it presents a minimum at $\theta =90^o$. For large $\eta $,
the opposite situation takes place and $\sigma _{Mott}$ has a maximum at 
$\theta =90^o$. An interesting situation occurs at the critical
value of the Sommerfeld parameter, $\eta _{C}$, where the cross section goes
through this transition. The value of $\eta _{C}$ is obtained from the
condition 
\begin{equation}
\left[ \frac{d^{2}\sigma _{Mott}(\theta )}{d\theta ^{2}}\right] _{\theta
=90^o}=0\,.  \label{criticond}
\end{equation}
Using eqs.(\ref{inc-coul}) and (\ref{int-coul}), we obtain 
\[
\left[ \frac{d^{2}\sigma _{Mott}(\theta )}{d\theta ^{2}}
\right]_{\theta =90^o}=16a^{2}\,\left[ \frac{1-2\eta ^{2}}{2s+1}+3\right] 
\]
and according to eq.(\ref{criticond}) we get 
\begin{equation}
\eta _{C}=\sqrt{3s+2}\,.  \label{etacrit}
\end{equation}

In figure 1, we show cross sections normalized to the value of the
Rutherford cross section at $90^o,$ $\sigma _{Ruth}(90^o)=a^2$, for collisions 
of identical bosons with spins $s=0$ and $s=1$. In
each case, the calculations were performed at $\eta _{C}$. I.e., 
$\eta =\sqrt{2}$ for $s=0$ and $\eta =\sqrt{5}$ for $s=1$. Also shown for
comparison is the incoherent cross section of eq.(\ref{inc-coul}),
normalized in the same way. Clearly, $\sigma _{inc}(90^o)/a^2=2$
and $\sigma _{b}(90^o)/a^2=4$ as shown in the figure. The
striking feature of the figure is the flatness of $\sigma _{Mott}$ over a
very wide angular region around $\theta =90^o$. It is essentially
constant for $60^o<\theta <120^o$, for $s=0$, and $70^o<\theta <110^o$, for $s=1$. 
This `transverse isotropy' (TI) is universal as the only relevant parameter which 
enters the discussion is the Sommerfeld parameter. In principle, it could be 
observed in atomic or nuclear systems at the appropriate energy. 
However, as we shall show below, the most appropriate case to investigate the above 
`transverse isotropy' is that of low-energy scattering of light identical nuclei, 
such us d-d or $\alpha -\alpha$.

Investigating a physical system which shows TI could shed light on several
small effects related to QED and possibly to QCD, as well as to atomic
effects in nuclear scattering. Also the assumed pure bosonic nature of the
multifermionic cluster could be nicely examined by a careful analysis of
data taken at $\eta _{C}$.

At this stage, it is important to investigate the optimal conditions for
observation of TI. The effective forces between identical nuclei or ionized
atoms are composed of the long range Coulomb part plus a shorter range
nuclear or Van der Waals force. Since the above discussion was based on a
pure Coulomb force, it is important to seek the physical conditions that
allows the neglect of the short range forces. Calling $E_{C}$ the collision
energy corresponding to $\eta _{C}$, the above condition corresponds to the
requirement that $E_{C}$ be sufficiently below the Coulomb barrier $V_{B}$,
the outermost maximum in the effective potential for $l=0$. In the collision
of identical particles of charge $q$ and mass $M$, the critical collision
energy $E_{C}$ is given by 
\[
E_{C}=\frac{Mq^{4}}{4\hbar ^{2}(3s+2)}\,. 
\]
If one approximates $V_{B}$ by the coulomb potential at the barrier radius 
$R_{B}$, namely 
\begin{equation}
V_{B}=\frac{q^{2}}{R_{B}}\,,  \label{VB}
\end{equation}
the condition $E_{C}<V_{B}$ yields 
\begin{equation}
\frac{Mq^{2}R_{B}}{\hbar ^{2}(3s+2)}<1\,.  \label{condition}
\end{equation}

Since the barrier radius in atomic collisions is very large, the above
condition cannot be satisfied. We then consider nuclear collisions. In the
nuclear physics one usually takes $R_{B}=1.4$ $\,(M/m_{0})^{1/3}\,$fm$,$
where $m_{0}$ is the nucleon mass. For light nuclei one can assume equal
number of protons and neutrons and set $M/m_{0}=2Z$, where $Z$ is the atomic
number. Eq.(\ref{condition}) then reduces to 
\begin{equation}
Z^{10/3}<25.4\,\times (2s+1)\,.  \label{condition1}
\end{equation}
It can immediately be checked that the above condition is only satisfied for 
$\alpha -$particles, in the case of $s=0$, and for $d-d$ and $^{6}Li-^{6}Li$
collisions, in the case of $s=1$ (if one relaxes the condition of equal
numbers of protons and neutrons, a couple of additional nuclei can be
included in this set). The Mott cross section at $\theta =90^o$ at
for collisions with the critical bombarding energy, $E_{C},$ is then given
by the simple expression 
\begin{equation}
\sigma _{Mott}(\theta =90^o)=\frac{\sigma _{0}}{Z^{6}}\,\left(
3s+2\right) ^{2}\,,  \label{sig90}
\end{equation}
with $\sigma _{0}=33.7$ barn.

In table I, we give the relevant quantities in the three above mentioned
cases. It is clear from this table that the transverse isotropy is an observable
phenomenon, although experimentally difficult in the case of $d-d$. An
important question to address now is the sensitivity of our result to the
uncertainty of the energy of the beam. Since a slight change in $\eta $\
form $\eta _{C}$ might wash out the TI, it is interesting to assess the
range of $\eta -$values around $\eta _{C}$\ which can still tolerate a
meaningfull study of the phenomenon. Figure 2 shows the way the shape of the
Mott cross section changes as $\eta $ is varied from $\eta _{C}$ by $5\%$.
This uncertainty in the $\eta -$value corresponds to about 10$\ \%$
uncertainty in the collision energy, which is certainly attainable in
existing accelerators. It is clear from figure 2 that the transverse
isotropy should be visible within a few per cent energy resolutions.

Although we have restricted our discussion so far to $E_{C}<V_{B}$, we
emphasize that the TI may come up at higher energies. However, in this case
the details of the short-range interaction becomes important and the
discussion becomes model- and system-dependent. We also point out that no
transverse isotropy is expected for Coulomb collisions of identical
fermions, since in this case both $\sigma _{inc}$ and $\sigma _{int}$ have
minima at $\theta =90^o$.  However, the situation may be different
when the short range interaction dominates.
We have briefly looked into this question by examining the extreme
case of hard sphere scattering with no Coulomb interaction.
The physical parameter that characterizes the collision process is $kR$ 
with $R$ being the sum of the radii of the two colliding particles.
Our preliminary results indeed show a TI for identical fermions with 
the rather large spin values $s \ge 9/2$. 
The critical value of $kR$ is the order of 2.5, which is of the same
order of magnitude as the value we found for bosons (1.5). 
More details of the present work with extensions to other potentials 
will be published elsewhere \cite{CDH}.

\bigskip\bigskip
This work was supported in part by CNPq and the MCT/FINEP/CNPq(PRONEX) under 
contract no. 41.96.0886.00.

\newpage

{\bf Tables}

\begin{itemize}
\item  Table I: The relevant quantities associated with the d-d, $^{6}Li-^{6}Li$ 
and $\alpha -\alpha$ collisions, at the critical value of the Sommerfeld parameter.
\end{itemize}

\begin{tabular}{p{1in}p{1in}p{1in}p{1in}p{1in}}
System & $s$ & $E_{C}($keV) & $V_{B}($keV) & $\sigma _{Mott}(90^o)\,($barn) \\ 
d + d & 1 & 5.0 & 400 & 135 \\ 
$^{6}$Li $+$\thinspace $^{6}$Li & 1 & 1200 & 2500 & 1.17 \\ 
$\alpha +\alpha $ & 0 & 400 & 1260 & 2.3
\end{tabular}

\bigskip
\bigskip
\bigskip

{\bf Figure Captions}

\begin{itemize}
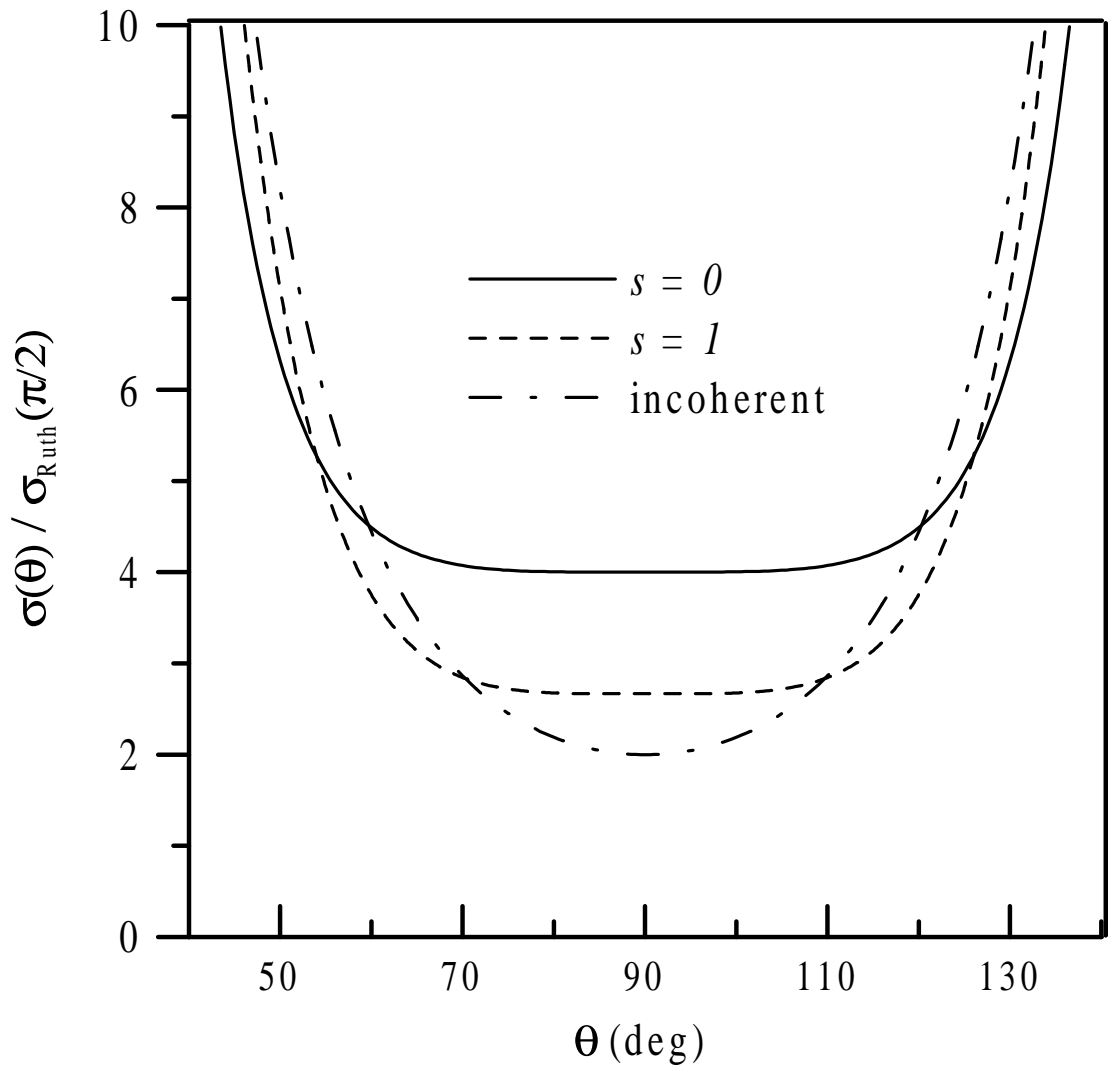
\item  Figure 1: The Mott cross sections for collisions of identical bosons
at the critical value of the Sommerfeld parameter. Results are shown in the
case of bosons with spin 0 (solid line) and spin 1 (dashed line), to which
corresponds respectively $\eta _{C}=\sqrt{2}$ and $\eta _{C}=\sqrt{5}$. For
comparison, the incoherent part of the cross section is also shown
(dot-dashed line).

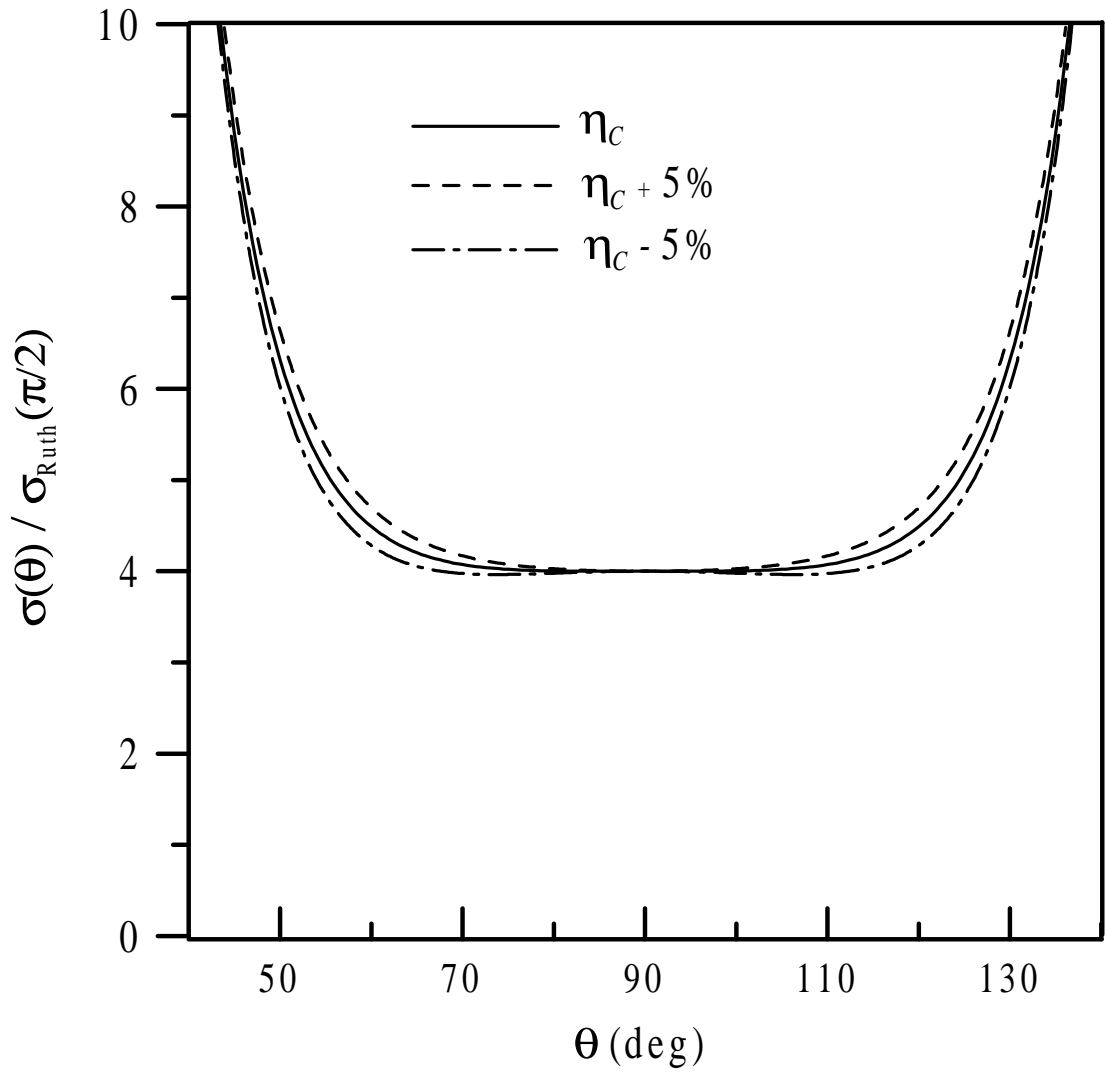
\item  Figure 2: Sensitivity of the transverse isotropy as $\eta $ deviates
from $\eta _{C}$ by 5 \%. The results are shown in the case of $s=0$.
\end{itemize}

\end{document}